\documentclass[aps,prc,10pt,twocolumn,superscriptaddress,showpacs,longbibliography]{revtex4-1}
\usepackage{slashbox}
\usepackage[x11names]{xcolor}
\colorlet{shadecolor}{LightBlue1}
\colorlet{framecolor}{Blue1}
\usepackage{graphicx}
\usepackage{amsmath}
\usepackage{verbatim}
\usepackage[utf8]{inputenc}
\usepackage{amsmath}
\usepackage{float}
\usepackage{cleveref}
\usepackage{xcolor}
\usepackage{soul}
\newcommand{\mev}{\, \text{MeV}}

\newcommand{\fm}{\, \text{fm}}
{\endMakeFramed}

\newcommand{\ket}[1]{\left| #1 \right>} 
\newcommand{\bra}[1]{\left< #1 \right|} 
\newcommand{\pilesseft}{\mbox{$\pi\text{\hspace{-5.5pt}/}$}EFT$\,$}

\newcommand{\cblack}{\color{black} }

\graphicspath{{./images_fin/}}
\begin{document}
 
\title{Tritium $\beta$-decay in pionless effective field theory}
\author{Hilla\ De-Leon} \email[E-mail:~]{hilla.deleon@mail.huji.ac.il}
\affiliation{Racah Institute of Physics, The Hebrew University of
 Jerusalem, 9190401 Jerusalem, Israel}
\author{Lucas\ Platter}
\email[E-mail:~]{lplatter@utk.edu} \affiliation{ Department of Physics
 and Astronomy, University of Tennessee Knoxville, TN 37996, USA}
\affiliation{Physics Division, Oak Ridge National Laboratory, Oak
 Ridge, TN 37831, USA}
 \author{Doron\ Gazit} \email[E-mail:~]{doron.gazit@mail.huji.ac.il}
 \affiliation{Racah Institute of Physics, The Hebrew University of Jerusalem,
 9190401 Jerusalem, Israel}

\date{\today}
\begin{abstract}
 We calculate the $\beta$-decay of tritium at next-to-leading order in pionless effective field theory. At this order, a low-energy parameter $l_{1, A}$ enters the calculation that is also relevant for a high-accuracy prediction of the solar proton-proton fusion rate. We use the tritium half-life to determine this parameter and provide uncertainty estimates. We show proper renormalization of our calculation analyzing the residual cutoff dependence of observables. We find that next-to-leading order corrections contribute about $4\%$ to the triton decay Gamow-Teller strength. We show that these conclusions are insensitive to different arrangements of the effective range expansion.
\end{abstract}
 \maketitle
 
\section{Introduction}
Weak decays of nuclei are an everyday window to quantum-chromodynamics
(QCD). This is the main reason for their extensive use in experiments
that try to study the limits of the Standard Model, from measuring the
masses of neutrinos using triton $\beta$-decay, to pin-pointing the
basic symmetries of the theory through the dynamics of $^6$He
decay. The triton $\beta$-decay, as the only $A=3$ $\beta$-decay, can,
therefore, probe unique properties of the nuclear force for both $^3$H
and $^3$He. Alas, since this reaction is characterized by its
low-energy, {\it i.e.,} $Q< 10 $ MeV, one cannot use QCD due to its
non-perturbative character at low energies.
 
In the past two decades, a novel theoretical method named effective
field theory (EFT) revolutionized nuclear physics. EFT is a simple,
order by order, renormalizable and model-independent theoretical
method that is used to describe low-energy processes. The prerequisite
for describing a physical process using EFT is that its transfer
momentum, $Q$, is small compared to a large scale, $\Lambda_{\rm cut}$
({\it i.e., } $Q/\Lambda_{\rm cut}\ll 1$) \cite{few, kaplan1,
 KSW1998_a, KSW1998_b,KSW_c} inherent to the system under
consideration. This method becomes particularly useful when there is a
significant scale separation between $Q$ and $\Lambda_{\rm cut}$, so
that only a small number of the effective operators corresponding to
the leading powers in $\frac{Q}{\Lambda_{\rm cut}}$ need to be
retained to reproduce long wavelength observables with the desired
accuracy.
 
The relevant momentum scale $Q$ is small, {\it i.e.}
$Q\ll\Lambda_{\rm cut}=m_\pi$ tritium $\beta$-decay and for many
other few-body electroweak processes of interest. In these cases,
pionless EFT (\pilesseft) is an appropriate
framework~\cite{Griesshammer_pionless, few_platter}. 
 
 
The Coulomb interaction in light nuclei is also an issue that needs to
be addressed. Na\"{i}vely, the Coulomb interaction is non-perturbative
at low momenta ($\lesssim 10\, \mev$) but should be perturbative in
nuclei where the typical momenta are much higher. However, recent
calculations done by K{\"o}nig et al. \cite{konig1,konig3,konig5} have
shown that for $A=3$, the Coulomb interaction can be treated
perturbatively, as reflected in the $^3$H-$^3$He binding energy
difference. Moreover, in Ref.~\cite{Big_paper}, we have shown that,
even though the Coulomb interaction does not conserve the
three-nucleon isospin, the Coulomb energy difference can be presented
in terms of a general matrix element between two $A=3$
bound-states. At LO, $^3$He (the lightest and, therefore, the simplest
nucleus that includes a Coulomb interaction) is described correctly
within \pilesseft, even when the Coulomb interaction is
non-perturbative \cite{3He,Quartet}. At next-to-leading order (NLO),
the results are not so clear, and some approaches point towards the
need for additional, isospin-dependent three-body forces. Therefore,
other three-nucleon observables are needed to obtain predictive power
within \pilesseft at NLO, such as the $^3$He binding energy
\cite{konig1, konig2, konig3}.
 
One way to test the predictive power of \pilesseft for the
three-nucleon system at NLO, and in particular, the effect of the
Coulomb interaction of such a system is through the aforementioned
electroweak properties of light nuclear systems. This is the goal of
this paper, in which we aim at describing tritium $\beta$-decay. This
observable is particularly interesting since it is well-known
experimentally, and can be used to determine the short-range strength
of the axial coupling to nuclei. Specifically, the \pilesseft
axial-vector current contains an additional two-body operator at NLO
whose low-energy constant (LEC), known as $l_{1, A}$ needs to be
determined. This is in particular important for a high-accuracy
description of the astrophysically relevant proton-proton fusion rate
\cite{pp_review,fermi_reference_1998,PhysRevLett.103.102502,PhysRevC.67.055206}.
However, its exact value is a matter of discussion in the literature
\cite{Kong1,Kong2,Butler:1999sv,Proton_Proton_Fifth_Order,L1A,Ando_proton}
and the lack of a \pilesseft calculation that is able to consistently determine
this counterterm is the reason for this discussion.

Besides determining $l_{1, A}$, we will also use our work to study the
convergence pattern of \pilesseft for this process and analyze the
residual cutoff dependence for this observable. Will also discuss the
impact of the Coulomb interaction on the relevant matrix element.
 
This paper is organized as follows: The general formalism of
\pilesseft is presented in Section~\ref{general_element}. The
\pilesseft formalism for the weak interaction is given in
Section~\ref{weak_interaction}. The general calculation of weak matrix
elements is presented in Section~\ref{matrix_element} while the
numerical results are given in \ref{numerical_results}. In
Section~\ref{extraction}, we use the experimental value of the triton
$\beta$-decay rate to fix $l_{1,A}$ at NLO. In
Section.~\ref{previous_L1A}, we compare this approach of matching this
counterterm to previous studies. We then summarize and provide an
outlook in Section~\ref{summary}.
 
\section{\pilesseft up to next-to-leading order}
\label{general_element}
Originally, \pilesseft was developed in terms of nucleon fields
alone. However, dynamical dibaryon fields provide a convenient way to
reformulate \pilesseft in a way that simplifies three-body
calculations. The fields $t$ and $s$ have quantum numbers of two
nucleons coupled to an S-wave spin-triplet and -singlet state,
respectively. The effective masses and interaction strengths of these
dibaryons are related to the two-nucleon scattering lengths and effective ranges in these two channels. This formulation is formally
equivalent to the usual single-nucleon theory but in a three-nucleon
calculation it reduces the problem to an effective two-body
calculation.

For the construction of the \pilesseft Lagrangian, we note that the
external momenta $q$ and the deuteron binding momentum $\gamma_t$ are
formally $\mathcal{O}(Q)$ ($Q$ is the typical momentum scale in the
reaction), the two-nucleon scattering lengths are $\mathcal{O}(1/Q)$,
and the two-nucleon effective range is
$\mathcal{O}({1/\Lambda_{\rm cut}})$. Up to NLO, {\it i.e.,}
$\mathcal{O}(Q/\Lambda_{\rm cut})$, the two-body Lagrangian has the
form \cite{rearrange}:
\begin{multline}
 \label{triton_Lagrangian}
 \mathcal{L}
 =N^{\dagger}\left (iD_0+\frac{{\bf D}^2}{2M}\right)N
 -t^{i\dagger}\left[\left (iD_0+\frac{{\bf D}^2}{4M}\right)-\sigma_t\right]t^i\\-
 s^{A\dagger}\left[\left (iD_0+\frac{{\bf
 D}^2}{4M}\right)-\sigma_s\right]s^A\\
 - y_t\left[t^{i\dagger}\left (N^TP_t^iN\right)+{\rm h.c.}\right]
 -
 y_s\left[s^{A\dagger}\left (N^TP_s^AN\right)+{\rm h.c.}\right]~,
\end{multline}
where $A$ denotes the isospin triplet index, $i$ the spin-triplet
index, and $N$ is the {\it single} nucleon field, of mass $M$ and
$P_t^i,\, P_s^A$ are the projection operators to the triplet and
singlet states, respectively. We note that the minus sign in front of
the kinetic terms of $s$ and $t$ implies that these fields are ghost
fields. The covariant derivative is:
\begin{equation}
 D_\mu = \partial_\mu + ie A_\mu \hat{Q}~,
\end{equation}
where $e$ is the electric charge and $\hat{Q}$ is the charge operator,
coupled to the electromagnetic field, $A_\mu$.
 
After renormalization one finds
$\sigma_{s}=\frac{2}{M\rho_{{s}}} \left(\frac{1}{a_{{s}}}-\mu
\right)$,
$\sigma_{t}=\frac{2}{M\rho_{{t}}} \left(\gamma_t-\mu \right)$, and
$y_{t,s}$ are the coupling constants between two nucleons and
dibaryon,
$y_{t,s} = \frac{ \sqrt{8\pi}}{M \sqrt{\rho_{{t,s}}}}\label{eq_y}$.
The expressions above contain the singlet scattering length $a_{s}$,
the deuteron binding momentum $\gamma_t$, and the triplet and singlet
effective ranges $\rho_{t}$ and $\rho_s$, respectively. The deuteron
binding momentum is related to the deuteron binding energy $B_2$
through $B_2 = \gamma^2/m$. The renormalization scale $\mu$ enters
these equations through the use of the power divergence subtraction
scheme \cite{KSW1998_b}. To include the Coulomb interaction, we will
also require the proton-proton ($pp$) scattering length $a_C$ and
effective range $\rho_C$. The experimental values for the two-nucleon
observables needed here are given in Tab. \ref{table_exp_data}.

In this work, contrary to previous works on the electroweak properties
of light nuclei, that set the renormalization group (RG) scale $\mu$
to $\mu=m_\pi$, we test correct renormalization by taking the UV
cutoff $\Lambda$ to infinity at the end of the calculation.
 
Na\"{i}vely, the effective ranges are fixed from scattering
experiments. However, since the triplet channel is bound, the deuteron
(spin-triplet, $t$) effective range can be alternatively fixed by the
long-range properties of the deuteron wave function.
 
The long-range properties of the deuteron wave function \cblack are set by a quantity that we will call $Z_d$, where $Z_d$ is defined through the deuteron asymptotic S-State normalization, ${A}_S$, such that ${A}_S\equiv\sqrt{2\gamma_t Z_d}$ and 
$Z_d = \frac{1}{1-\gamma_t \rho_t} \approx 1.690(3)$ \cite{Phillips:1999hh} \cblack ($\gamma_t$ is
the deuteron binding momentum, $\gamma_t=\sqrt{M E_b(d)}$). In the
effective range expansion (ERE), the order-by-order expansion of $Z_d$
is:
\begin{equation}\label{eq_Zd_Q}
 \begin{split}
 Z_d^{\text{LO}}&=1~,\\
 Z_d^{\text{NLO}}&=1+\gamma_t\rho_t \approx 1.408\\
 \end{split}
\end{equation}

\begin{table}[H]
 \begin{center}
 \begin{tabular}{ccc}
 \centering
 Parameter& Value & Reference \\
 \hline
 $\gamma_t$& 45.701 MeV &\cite{32} \\ 
 $\rho_t$& 1.765 fm &\cite{deSwart:1995ui}\\
 $a_s$& -23.714 fm &\cite{Preston_1975}\\ 
 $\rho_s$& 2.73 fm &\cite{deSwart:1995ui}\\
 $a_p$& -7.8063$\pm$0.0026 fm &\cite{34} \\
 $\rho_C$& 2.794$\pm$0.014 fm &\cite{34}
 \end{tabular}
 \caption{\footnotesize{Parameters used in the numerical calculation}}
 \label{table_exp_data}
 \end{center}
\end{table}
This result for the perturbative expansion of the Z-factor is based on
a matching of the parameters in the EFT to the effective range
expansion (ERE). At NLO, the parameters can also be chosen to fix the
pole position and wave function renormalization constant of the
triplet two-body propagator to the deuteron values. This
parameterization is known as the Z-parameterization and is
advantageous because it reproduces the correct residue about the
deuteron pole at NLO, instead of being approached perturbatively order
by-order as in ERE-parameterization
\cite{Griesshammer_3body,Kong2,Vanasse,Vanasse:2015fph,Phillips:1999hh}:
 
\begin{equation}\label{eq_Zd_2}
 \begin{split}
 Z_d^{\text{LO}}&=1~,\\
 Z_d^{\text{NLO}}&=1+\left(Z_d^{\text{full}}-1\right) = 1.690(3).
 \end{split}
\end{equation}
The price is that the value of the triplet effective range at NLO in
this parameterization is
$\rho'_t=\frac{Z_d-1}{\gamma_t}\approx
\frac{0.690}{\gamma_t}=2.979\fm$.
In the following, we use both parameterizations at NLO.

\subsection{$A=3$ nuclear amplitudes, matrix elements, and regularization}
While an analytical result for the deuteron wave function can be
derived in pionless EFT~\cite{Kong1}, the three-nucleon scattering
amplitude has to be calculated numerically. The different channels for
$^3$H are the spin-triplet - $t$ (representing an ``off-shell''
deuteron, $d$, dibaryon), and the spin-singlet - $s$ ($nn,np$). For
$^3$He, the contributing channels are the spin-triplet - $t$,
spin-singlet - $s$ ($np$) and $pp$ \cite{faddeev}. The latter is
required because of the Coulomb force between the protons, which
modifies the long-range scattering properties of these nucleons.
 
The Faddeev integral equation, used in this EFT, has to be
regularized. A simple way to do this is to evaluate the momentum space
integrals in the integral equation up to a cutoff $\Lambda$. Since
\pilesseft is supposed to be order-by-order renormalizable; the theory
should not depend on this ultraviolet (UV) cutoff. However, for
three-nucleon systems, the numerical and theoretical solution of the
integral equations displays a strong dependence on the cutoff. To
overcome this problem, one adds a three-body counterterm
\cite{3bosons,triton}. In the case of $^3$He, the addition of Coulomb
interaction to the three-nucleon amplitude leads to divergence in the
Coulomb Feynman diagrams, which is solved by the redefinition of the
proton-proton scattering length
\cite{Coulomb_effects,proton_proton_scattring}. With this
redefinition, the $^3$He binding energy is renormalization-group
invariant at LO. \cite{konig3,konig5}.
 
The calculation of bound-state amplitudes requires the solution of a
homogeneous Faddeev equation defined up to a normalization. The
calculation of next-to-leading order corrections follows the same
formalism, however with a single NLO insertion, e.g., of an effective range. The three-body wave function normalizations,
$Z_{^3\text{H}},\,Z_{^3\text{He}}$, are calculated diagrammatically,
by summing over all possible connections between two identical vertex
functions as presented in Ref.~\cite{Big_paper}.
 
The Coulomb force is included by considering the full $pp$ Coulomb
propagator and allowing a single photon insertion in the three-body diagrams. At LO, it was shown that $^3$He is described correctly
within \pilesseft~\cite{Quartet,3He,2016PhLB..755..253K}, while at
NLO, within the power counting considered here, there is a need for an
additional, isospin dependent, three-body force to renormalize the
$^3$He binding energy ~\cite{konig1,konig2,konig3,Big_paper}.
 
The nuclear amplitudes we use here are taken explicitly from
Ref.~\cite{Big_paper}, where they were benchmarked numerically, and
validated using the binding energy difference between $^3$H-$^3$He.
 
\section{The weak interaction in \pilesseft }
\label{weak_interaction}
For low-energy charge lowering processes, the weak-interaction Hamiltonian is
\begin{equation}\label{lweak}
 \mathcal{H}_{\text{Weak}}=\frac{G_FV_{ud}}{\sqrt{2}}l_+^{\mu }J_{\mu
 }^-,
\end{equation}
where $G_F$ is the Fermi constant and $V_{ud}$ is the CKM matrix
element. $l^\mu$ is the lepton current and $J_\mu$ is the hadronic
current. We use the two-body hadronic current $J_\mu$ from the
\pilesseft effective Lagrangian with dibaryon fields up to
NLO. 
 
The hadronic current contains two parts, a polar-vector and
axial-vector, $J_\mu=V_\mu-A_\mu$. The part of the polar vector
current relevant to $\beta$-decay with a vanishing energy transfer is:
\begin{equation}
 \label{V}
 V_0^{\pm}=N^{\dagger}\frac{{\tau^\pm}}{2}N~, 
\end{equation}
where $\tau^{\pm}=\tau_1\pm i\tau_2$.
 
Here, we utilized the fact that the Conserved Vector Current (CVC)
hypothesis is accurate at this order of EFT.
 
The axial-vector part is (see Ref.~\cite{Ando_proton,rearrange}): 
\begin{equation}
 \label{eq_weak_axial2}
 A^{\pm}_i=\underbrace{\frac{g_A}{2}N^\dagger
 \sigma_i\tau^{\pm}N}_{\text{LO}}
 +\underbrace{g_AL_{1, A} \left (t_i^\dagger s_{\pm}+{\rm h.c.}\right)}_{\text{NLO}}~,
\end{equation}
where $\tau^\pm = \tau_1 \pm i \tau_2$ and $s_\pm = s_1 \pm i s_2$ and
$g_{A}$ is the axial coupling constant for a single nucleon, known
from neutron $\beta$-decay. We denote the coefficient of the two-body
operator with $L_{1,A}$ (see Ref.~\cite{Big_paper} for more
details). A number of previous pionless EFT electroweak calculations
were calculated using a Lagrangian with single nucleon fields
only. The axial-vector two-body counterterm takes then the form
\begin{equation}
 \label{eq:L1Asingle-nucleon}
 L'_{1,A} (N^T P_i N)^\dagger (N^T P_{-} N)~.
\end{equation}
The coefficients of these two-body operators are related through the relation
\begin{equation}\label{eq_L1A}
 {L_{1, A}}(\mu)=-\frac{\rho_t+\rho_s}{2\sqrt{\rho_s\rho_t}}+
 \frac{L'_{1, A}}{2\pi g_A}\frac{1}{\sqrt{\rho_s\rho_t}}\left(\mu-\gamma_t\right)
 \left(\mu-\frac{1}{a_s}\right)~.
\end{equation}
It was already pointed out by Kong and Ravndal~\cite{Kong2} that
renormalization scale dependence of $L'_{1,A}$ can be made explicit by
writing
\begin{equation}
 \label{eq:l1a-relation1}
 l_{1,A}=
 \frac{L'_{1, A}}{2\pi g_A}\frac{1}{\sqrt{\rho_s\rho_t}}
 \left(\mu-\gamma_t\right)\left(\mu-\frac{1}{a_s}\right)~,
\end{equation}
where $l_{1,A}$ is now a cutoff (renormalization scale) independent
number that characterizes the physics underlying the coupling of the
external axial current to two-nucleon system.

In two-nucleon calculations the renormalization scale $\mu$ was
frequently set to the breakdown scale of theory, usually at $m_\pi$.
In this work, $\mu$, the renormalization scale, is set to the momentum
space cutoff employed in the three-nucleon integral equations, 
$\mu=\Lambda$.

We will show below that by taking $\Lambda\rightarrow \infty$
numerically that when we use these relations we can obtain results for
the constant $l_{1, A}$ that are converged with respect to the cutoff
$\Lambda$.
 
\section{$^3$H $\beta$-decay matrix elements}
\label{matrix_element}
In this section, we outline the calculation of the matrix element of
the weak reaction:
\begin{equation}\label{eq_beta}
 ^3\text{H}\rightarrow ^3\text{He}+e^-+\overline{\nu_e}~. 
\end{equation}
This $\beta$-decay matrix element can be calculated using the LO and
NLO $A=3$ bound-state wave functions, as introduced in
Ref.~\cite{Big_paper}.

\subsection{$^3$H $\beta$-decay observables}
The half-life of $^3$H $\beta$-decay can be expressed as~\cite{fermi_reference_1998}:
\begin{equation}\label{eq_GT}
 fT_{1/2}=\frac{K/G_V^2}{\langle\| \text{F} \|\rangle^2+g_A^2 \frac{f_A}{f_V}\langle ||\text{GT}|| \rangle^2}~,
\end{equation}
where $ fT_{1/2}=1129.6\pm 3$~\cite{t_reference} is the triton
comparative half-life, $K=2\pi^3\log2/m_e^5$ (with $m_e$ denoting the
electron mass), $G_V$ is the weak interaction vector coupling constant
(such that $K/G_V^2=6146.66 \pm 0.6$ \cite{K_reference}),
$f_V=2.8355 \times 10^{-6}$ and $f_A=2.8506 \times 10^{-6}$ are the
Fermi functions calculated by Towner, as reported by Simpson in Ref.
~\cite{fv_reference}. $\langle\| \text{F} \rangle\|$ and
$\langle \|\text{GT}\| \rangle$ are the reduced matrix elements of the vector
and axial current $A=3$ wave-function, respectively.

\subsection {General $A=3$ matrix element in \pilesseft}
\label{general_matrix}
The weak transitions
$\langle \|\text{GT}\| \rangle,\ \langle \|\text{F}\| \rangle$ are defined as
matrix elements between the initial state wave function
$\psi^{^3\text{H}}$, and the final state, $\psi^{^3\text{He}}$, using
the general mechanism introduced in Ref \cite{Big_paper}.
 
\subsubsection{$A=3$ one-body matrix element}
In Ref.~\cite{Big_paper}, we showed that at LO, the three-nucleon
normalization can be written as:
\begin{equation}
 1= \sum\limits_{\mu,\nu}
 \bra{\psi^{i}_\mu}
 \mathcal{O}_{\mu\nu}^{\text{norm}}(E_i)\ket{\psi_\nu^{i}}~,
 \end{equation}
 where $\mathcal{O}_{\mu\nu}^{\text{norm}}(E_i)$ is the normalization operator such that: 
 \begin{multline}\label{eq:1b:Onorm}
 \mathcal{O}_{\mu\nu}^{\text{norm}}(E_i)=\\
 \frac{\partial}{\partial_E}\left [\hat{I}_{\mu\nu}(E,p,p')-My_\mu y_\nu a^i_{\mu\nu}K_{\mu\nu}^i(p',p,E)\right]\bigg|_{E=E_i}~,
 \end{multline}
 where:
 \begin{equation}
 K_{\mu\nu}^i=\begin{cases}
 K_0(p',p,E)&i=^3\text{H}\\
 K_0(p',p,E)+K_{\mu\nu}^C(p',p,E)&i=^3\text{He}
 \end{cases}~,
 \end{equation}
 and
 \begin{eqnarray}
 \hat{I}_{\mu\nu}(p,p',E)&=&\frac{2\pi^2}{{ p}^2}\delta\left(p-p'\right)D_{\mu}(E,p)^{-1}\delta_{\mu,\nu}\\
 K_0 (p,p',E)&=&\frac{1}{2pp'}Q_0\left (\frac{p^2+p'^2-ME}{pp'}\right)~,
 \end{eqnarray}
 where $\delta_{\mu,\nu}$ is the Kronecker delta and:
 \begin{gather}
 \label{Q}
 Q_0 (\text{a})=\frac{1}{2} \int^1_{-1}\frac{1}{x+a}dx~.
 \end{gather}
 \cblack

 Since we will consider a matrix element of triton and Helium-3,
 it is convenient to express the wave of the triton in terms of three
 components $t,np$ and $nn$. Note that we assume here that
 $a_{nn}=a_{np}=a_s$. The coefficients $a_{\mu\nu}$ are then
 \begin{equation}
 \label{eq_a_mu_nu_3H}
 a^{^3\text{H}}_{\mu\nu}=a_{\mu\nu}=\begin{array}{c|ccc}
 \mbox{\backslashbox{$\nu$\kern-1em}{\kern-1em$\mu$}}&t&np&nn \\ \hline
 t&-1&3&3\\
 np&1&1&-1\\
 nn&2&-2&0
 \end{array}~,
 \end{equation}
 and 
 \begin{equation}
 \label{eq_a_mu_nu_3He}
 a^{^3\text{He}}_{\mu\nu}=a'_{\mu\nu}=\begin{array}{c|ccc}
 \mbox{\backslashbox{$\nu$\kern-1em}{\kern-1em$\mu$}}&t&np&pp \\ \hline
 t&-1&3&3\\
 np&1&1&-1\\
 pp&2&-2&0
 \end{array}~,
 \end{equation}
$K_{\mu\nu}^C(p'',p',E)$ is the $\mu,\nu$ index of the one-photon
 exchange matrix, $K^C (p'', p', E)$ (see Ref.~\cite{Big_paper}),
 $\mu =t,np,nn$ are the different triton channels, $\mu =t,s,pp$ are the
 different $^3$He channels, $y_{\mu,\nu}$ are the nucleon-dibaryon
 coupling constants for the different channels, $a_{\mu\nu}$
 ($a'_{\mu\nu}$) are a result of the $n-d$ ($p-d$) doublet-channel
 projection (\cite{Parity-violating}) and $D_{\mu}(E,p)$ is the
 dibaryon propagator (\cite{Big_paper,triton,3bosons}).
 
 A general one-body operator, can be written as a generalization of a
 three-nucleon normalization operator for the case of both energy and
 momentum transfer, between initial (i) and final (j) $A=3$ bound-state
 wave functions ($\psi_{i,j})$. The general operator
 $\mathcal{O}_{j,i}$ factorizes into the following parts:
 \begin{equation}
 \mathcal{O}_{j,i}=\mathcal{O}^{J}\mathcal{O}^{T}\mathcal{O}_{j,i}(q_0,q), 
 \end{equation}
 where $\mathcal{O}^{J}$, the spin part of the operator whose total
 spin is $J$, and $\mathcal{O}^{T}$, the isospin part of the operator,
 that depend on the initial and final quantum numbers. The spatial part
 of the operator, $\mathcal{O}_{j,i}(q_0,q)$, is a function of the
 three-nucleon wave function's binding energies, ($E_i$,$E_j$) and the
 energy and momentum transfer ($q_0,q$, respectively).
 
 In the case of triton $\beta$-decay, the spin and isospin one-body
 operators are combinations of Pauli matrices, so their reduced matrix
 elements $\langle \| \text{F} \|\rangle$ and $\langle \| \text{GT}\|\rangle$ can be
 easily calculated as a function of the three-nucleon quantum total
 spin and isospin numbers. In Ref.~\cite{Big_paper} we showed that the
 reduced matrix element of such an operator can be written as:
 \begin{multline}\label{eq_general_operator_reduced}
 \langle\| \mathcal{O}_{j,i}^{\text{1B}}(q_0,q)\|\rangle=
 \left\langle\frac{1}{2}\left\|\mathcal{O}^J\right\|\frac{1}{2}\right\rangle 
 \\
 \times
 \sum\limits_{\mu,\nu}
 \bra{\psi^j_\mu}y_\mu y_\nu\Bigl\{d'^{ij}_{\mu\nu} \hat{\mathcal{I}}(q_0,q)\\+
 a'^{ij}_{\mu\nu}\left [\hat{\mathcal{K}}(p,p',E,q_0)+{\hat{\mathcal{K}}^C_{\mu\nu}}(q_0,q)\right]\Bigr\} \ket{\psi^i_\nu} 
 ~,
 \end{multline}
 such that for $i=j$:
 \begin{eqnarray}
 d'^{ii}_{\mu\nu}&=&\delta_{\mu,\nu}\\
 a'^{ii}_{\mu\nu}&=&\begin{cases}
 a_{\mu\nu}&i=j=^3\text{H}\\
 a'_{\mu\nu}&i=j=^3\text{He}\\
 \end{cases}.
 \end{eqnarray}
 The spatial parts of the operator are denoted by
 $\hat{\mathcal{I}}(E, q_0,q)$, $\hat{\mathcal{K}}(q_0,q)$ and
 ${\hat{\mathcal{K}}^C_{\mu\nu}}(E, q_0,q)$. The full analytical
 expressions for $\hat{\mathcal{I}}(E, q_0,q)$ and
 $\hat{\mathcal{K}}(E, q_0,q)$ are given in Ref.~\cite{Big_paper} while
 ${\hat{\mathcal{K}}^C_{\mu\nu}}(E, q_0,q)$ are the diagrams that
 contain a one-photon interaction in addition to the energy and
 momentum transfer. A derivation of an analytical expression for these
 diagrams is too complex, so they were calculated numerically
 only. $a'^{ij}_{\mu\nu}$ and $d'^{ij}_{\mu\nu}$ are a result of the
 $N-d$ doublet-channel projection coupled to
 $\mathcal{O}^J\mathcal{O}^T$ (for more details, see
 Ref~\cite{Big_paper}).
 \subsubsection{Two-body matrix element}
 In contrast to the normalization operator given in Eq.~\eqref{eq:1b:Onorm},
 which contains only one-body interactions, a typical \pilesseft
 electroweak interaction contains also the following two-body
 interactions up to NLO:
 \begin{equation}
 t^\dagger t,\,s^\dagger s,\,(s^\dagger t+h.c)~,
 \end{equation}
 under the assumption of energy and momentum conservation. 
 The diagrammatic form of the different two-body interactions is given in Ref.~\cite{Big_paper}.
 
 \subsection{Fermi and Gamow-Teller matrix elements}
 The Gamow-Teller operator of the triton $\beta$-decay
 ($\langle ||\text{GT}|| \rangle$) matrix element is given by:
 \begin{multline}
 \label{Eq_GT}
 \langle\| \text{GT}\|\rangle=\frac{\langle\psi^{^3\text{He}}\|\boldsymbol{A}^{+}\|\psi^{^3\text{H}}\rangle}{g_A\sqrt{2}}=
 \\
 {\Bigl\langle\frac{1}{2}\Bigl\|\tau^{+}\Bigr\|\frac{1}{2}\Bigr\rangle}\dfrac{\langle\frac{1}{2}\|\boldsymbol{\sigma}\|\frac{1}{2}\rangle}{\sqrt{2}}
 \\
 \times\sum\limits_{\mu,\nu}
 {\psi^{^3\text{He}}_\mu(p')\otimes}y_\mu y_\nu\Bigl\{d'^{ij}_{\mu\nu} \hat{\mathcal{I}}(q_0,q)
 \\
 + a'^{ij}_{\mu\nu}\left[\hat{\mathcal{K}}(q_0,q)+\hat{\mathcal{K}}_{\mu\nu}^C(q_0,q)\right]\Bigr\} \otimes{\psi^{^3\text{H}}_\nu(p)}
 \\
 -L_{1, A}\left(\frac{2}{3}\langle\psi^{^3\text{H}}_{nn}|\psi^{^3\text{He}}_t\rangle+\langle\psi^{^3\text{H}}_t|\psi^{^3\text{He}}_{pp}\rangle\right)~,
 \end{multline} 
 where:
 \begin{equation}\label{eq_otimes}
 A (..., p)\otimes B (p, ...)=\int A (.., p)B (p, ...)\frac{p^2}{2\pi^2}dp~.
 \end{equation}
 
 $d'^{ij}_{\mu\nu}$ is given by:
 \cblack
 \begin{equation}
 d'_{\mu\nu}=\begin{array}{c|ccc}
 \mbox{\backslashbox{$\nu$\kern-1em}{\kern-1em$\mu$}}&t&np&pp \\
 \hline
 t&1/3&0&-1\\
 np&0&1/3&0\\
 nn&-2/3&0&0
 \end{array}
 \end{equation}
 and \begin{equation}
 \mbox {$a'^{ij}_{\mu\nu}$}=\begin{array}{c|ccc}
 \mbox{\backslashbox{$\nu$\kern-1em}{\kern-1em$\mu$}}&t&np&pp \\ \hline
 t&-7/3&1&3\\
 np&1&1&-1\\
 nn&2/3&-2&-2
 \end{array}
 \end{equation}
 \cblack
 
 The reduced Fermi matrix element $\langle|| \text{F}|| \rangle$ is given by
 \begin{multline}\label{Eq_F}
 \langle\| \text{F} \|\rangle=\frac{\langle\psi^{^3\text{He}}\|V^+\|\psi^{^3\text{H}}\rangle}{\sqrt{2}}=
 \\{\Bigl\langle\frac{1}{2}\Bigl\|\tau^+\Bigr\|\frac{1}{2}\Bigr\rangle}\sum\limits_{\mu,\nu}
 {\psi^{^3\text{He}}_\mu(p')\otimes}y_\mu y_\nu\Bigl\{d'^{ij}_{\mu\nu} \hat{\mathcal{I}}(q_0,q)
 \\+ a'^{ij}_{\mu\nu}\left[\hat{\mathcal{K}}(q_0,q)+\hat{\mathcal{K}}_{\mu\nu}^C(q_0,q)\right]\Bigr\} \otimes{\psi^{^3\text{H}}_\nu(p)}~,
 \end{multline}
 where \begin{equation}\label{eq_d_fermi}
 d'^{ij}_{\mu\nu}=\begin{array}{c|ccc}
 \mbox{\backslashbox{$\nu$\kern-1em}{\kern-1em$\mu$}}&t&np&pp \\
 \hline
 t&1&0&0\\
 np&0&1&-1\\
 nn&0&2&0
 \end{array}
 \end{equation}
 and \begin{equation}\label{eq_a_fermi}
 \mbox{ $a'^{ij}_{\mu\nu}$}=\begin{array}{c|ccc}
 \mbox{\backslashbox{$\nu$\kern-1em}{\kern-1em$\mu$}}&t&np&pp \\ \hline
 t&-1&3&3\\
 np&1&1&1\\
 nn&2&-2&-2
 \end{array}~,
 \end{equation}
 $\mu, \nu$ denote the different channels of the three-nucleon wave
 function ($t, np, pp$ for $^3$He and $t, np,nn$ for $^3$H), where
 $\psi_{\mu}, \psi_{\nu}$ are the three-nucleon wave functions for the
 different channels, defined using the homogeneous solution of the
 three-nucleon scattering amplitude \cite{Big_paper} and
 $q_0=E_{^3\text{He}}-E_{^3\text{H}}$ is the energy transfer.
 
\begin{figure}[h!]
 \begin{center}
 \includegraphics[width=1\linewidth]{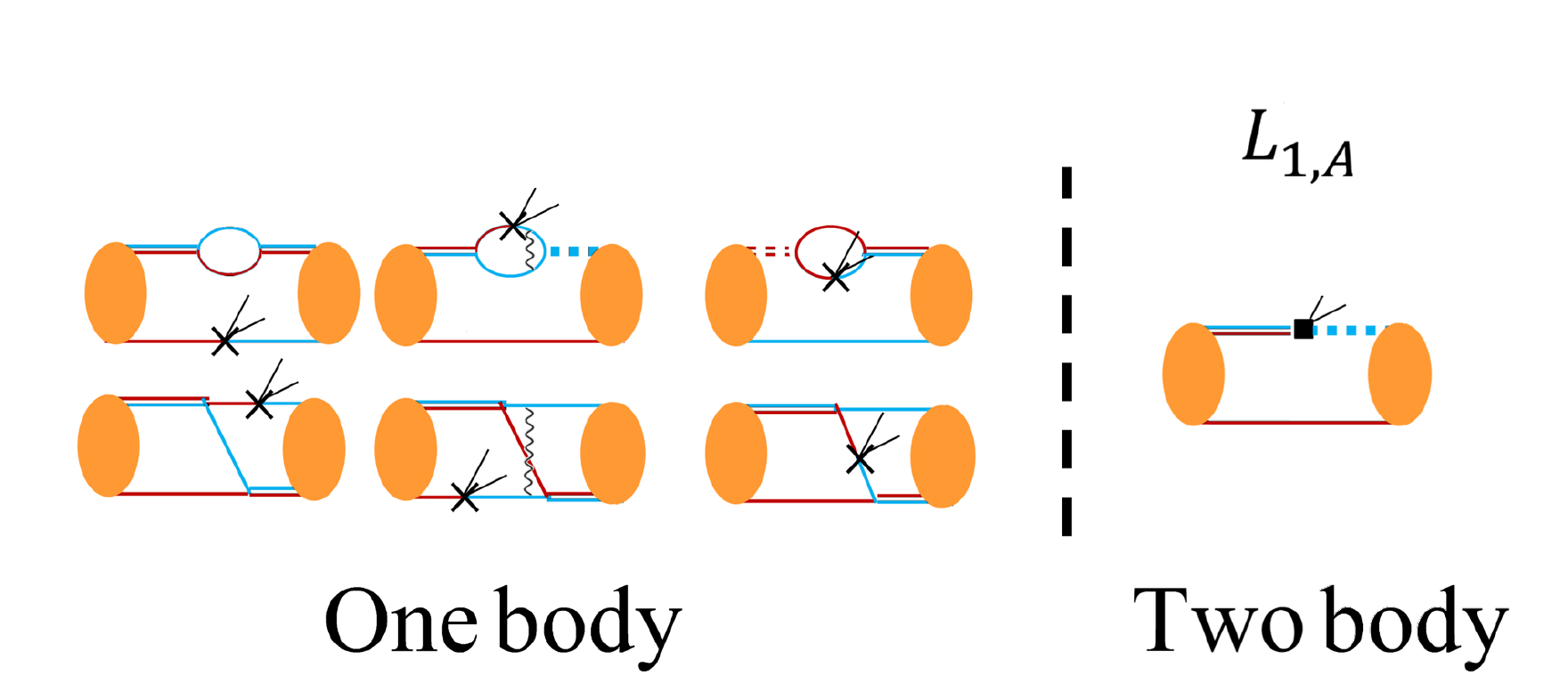}\\
 \caption{\label{fig_weak_topo}Different topologies of the diagrams
 contributing to the triton $\beta$-decay amplitude. The LHS of each
 diagram is $^3$H, while the RHS is $^3$He. The double lines are the
 propagators of the two dibaryon fields $D_t$ (solid), $D_s$ (dashed
 for $nn$ and $np$, dotted for $pp$), where the red (blue) lines
 denote a neutron (proton) propagator. Most of the diagrams couple
 both the triplet and the singlet channels. The diagrams with
 one-body interactions contribute to both
 $\langle ||\text{F} ||\rangle$ and $\langle ||\text{GT}||\rangle$
 transitions, \cblack while the two-body interactions are coupled to
 the effective ranges $\rho_t$ and $\rho_s$ and to $l_{1, A}$ ( and
 contribute only for the GT transition, where $L_{1,A}$ is defined
 in Eq.~\eqref{eq_L1A}).}
 \end{center}
 \end{figure}

 The general diagrammatic form of $^3$H $\beta$-decay, shown in
 Fig.~\ref{fig_weak_topo}, is similar to the general matrix element
 introduced in Ref.~\cite{Big_paper}. For both Fermi and
 GT transitions, the left-hand side (LHS) bubbles of
 the diagrams are $^3$H, while the right-hand side (RHS) bubbles are
 $^3$He.
 
 The one-body diagrams that contain a one-body weak interaction
 contribute to both $\langle ||\text{F} ||\rangle$ and
 $\langle ||\text{GT}||\rangle$ transitions. These one-body diagrams
 are taken up to NLO, and, therefore, contain the NLO insertion for
 the one-body diagrams, as discussed in Ref.~\cite{Big_paper}. The
 two-body diagrams include a two-body term originating from the ERE
 term in the Lagrangian,
 $\frac{1}{2}\frac{\rho_t+\rho_s}{\sqrt{\rho_t\rho_s}}g_A$, and the
 two-body operator whose coefficient is proportional to $l_{1,
 A}$. These diagrams contribute to the GT transition only.
 \section{Numerical results}\label{numerical_results}
 \subsection{Fermi operator}
 In the absence of the Coulomb interaction, $^3$H is identical to
 $^3$He and the Fermi transition is equal to the triton wave function
 normalization as defined in Ref.~\cite{Big_paper}:
 \begin{multline}\label{Eq_F_triton}
 \langle ||\text{F}|| \rangle^{0}=\frac{\langle\psi^{^3\text{H}}\|{\tau}^0\|\psi^{^3\text{H}}\rangle}{\sqrt{2}}={\Bigl\langle\frac{1}{2}\Bigl\|\tau^0\Bigr\|\frac{1}{2}\Bigr\rangle}\\
 \times\sum\limits_{\mu,\nu}
 {y_\mu y_\nu\psi^{^3\text{H}}_\mu(p')\otimes}\left[d'^{ii}_{\mu\nu} \hat{\mathcal{I}}(0,0)\right.\\
\left. +a'^{ii}_{\mu\nu}\hat{\mathcal{K}}(0,0)\right] \otimes{\psi^{^3\text{H}}_\nu(p)}=1~,
 \end{multline}
 where in the absence of the Coulomb interaction: 
 \begin{align}
 d'^{ii}_{\mu,\nu}&=\delta_{\mu,\nu}\label{eq_d_3He} \\
 a'^{ii}_{\mu\nu}&=a_{\mu\nu}\label{eq_a_3He}
 \end{align}
 
 From comparison between \cref{eq_a_mu_nu_3H,eq_a_mu_nu_3He} with
 \cref{eq_d_fermi,eq_a_fermi}, we expect that
 $\langle ||\text{F}|| \rangle=1-\epsilon$ \cite{fermi_reference_1998},
 where $\epsilon\ll1$ originates mostly from the isospin breaking due
 to the Coulomb interaction. We can, therefore, examine the effects
 of isospin breaking on the Fermi transition due to the Coulomb
 interaction, and the additional one-photon exchange diagrams and then
 compare them to the Gamow-Teller transition. In this section, we
 present our calculations of the Fermi transition. First, we calculate
 the Fermi transition in the absence of the Coulomb interaction but
 under the assumption that $a_{nn, np}\neq a_{pp}$. Second, we
 calculate the Fermi transition with $\alpha \neq 0$, and, obviously,
 $a_{nn, np}\neq a_{pp}$, as a result, for both Z- and
 ERE-parameterization. All these calculations result from the LHS of
 the diagrams in Fig.~\ref{fig_weak_topo}.
 
\begin{figure}[h!]
 \centerline{
 \includegraphics[width=1\linewidth]{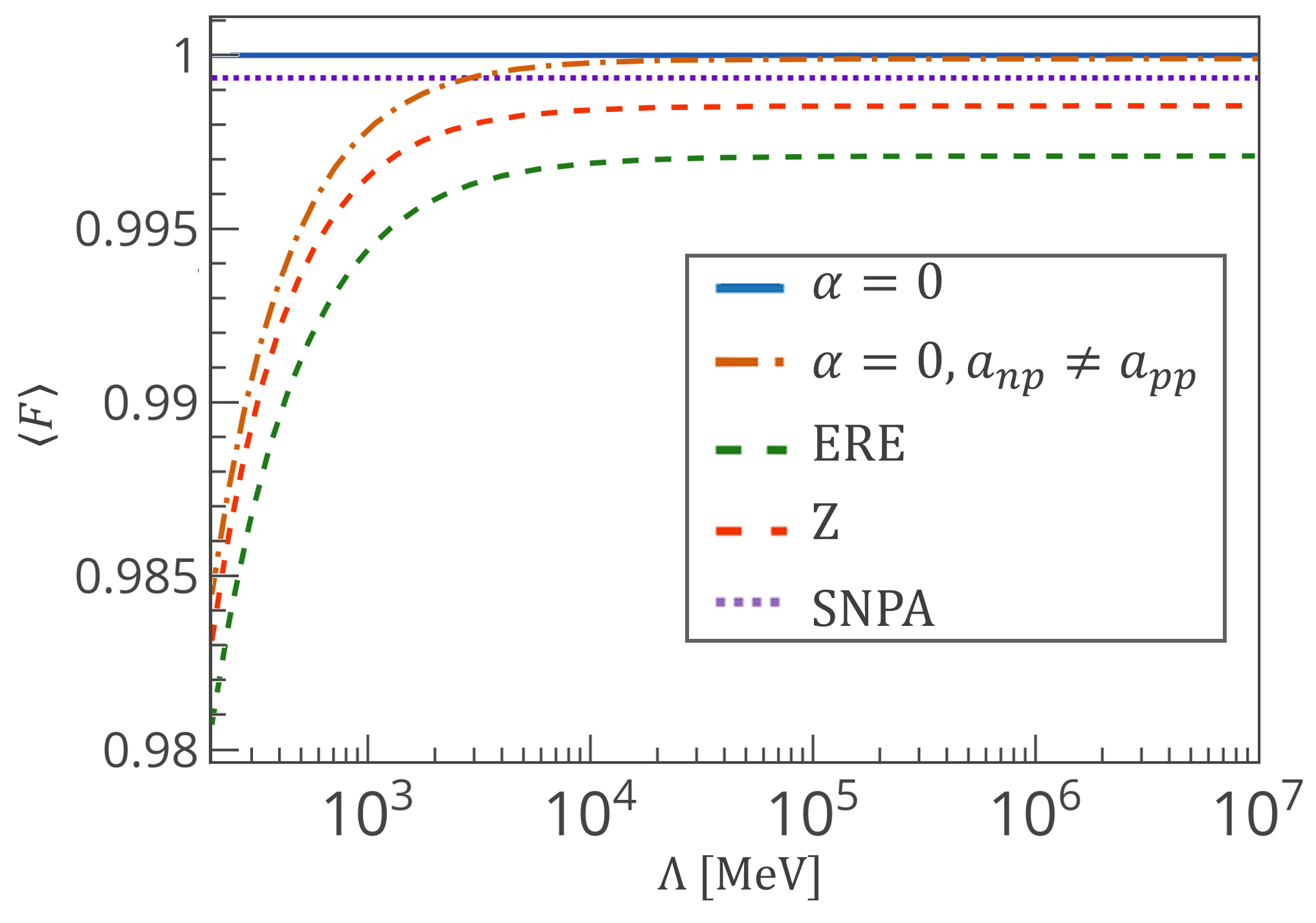}}
 \caption{\label{fig_tauminus}\footnotesize{Numerical results
 of the Fermi transition. The solid line is the LO result of
 $\langle ||\text{F}|| \rangle=1$ with $\alpha=0$. The dashed-dotted
 line is the numerical result for $\alpha=0$ with isospin
 breaking effects in the scattering length: $a_{np}\neq a_{pp}$
 (LO). The green (red) dashed line shows the numerical result at LO for the ERE- (Z-) parameterization 
 with $\alpha\neq0$. The dotted line is plotted at the value of
 $\langle ||\text{F}|| \rangle=0.9993$ from Ref.~\cite{fermi_reference_1998}}. 
}
 \end{figure}

 We use the experimental data shown in Tab.~\ref{table_exp_data} as
 input for our numerical calculations shown in Fig.~\ref{fig_tauminus}
 and Tab.~\ref{table_F}.
 
 \begin{table}[H]
 \begin{center}
 \begin{tabular}{l |c}
 \centering
 &$\langle ||\text{F}|| \rangle$\\
 \hline
 One-body, LO $\alpha=0$& 1\\
 One-body, LO $\alpha=0, a_{np} \neq a_{pp}$& 0.9999\\
 LO, ERE &0.9971\\
 LO, Z &0.9985\\
 SNPA \cite{fermi_reference_1998}&0.9993
 \end{tabular}
 \caption{\footnotesize{Numerical results of
 $\langle ||\text{F}|| \rangle$. Note that the second row is
 without an explicit Coulomb force ($\alpha=0$) but with isospin
 breaking in the scattering lengths, {\it i.e.,} with the physical
 values for the scattering lengths $a_{np} \neq a_{pp}$.}}
 \label{table_F}
 \end{center}
 \end{table}

 Our numerical result compares well to the $\langle ||\text{F}|| \rangle$ standard
 nuclear physics approach (SNPA) calculation by Schiavilla {\it et
 al.}~ \cite{fermi_reference_1998}. The SNPA calculation involves
 nuclear wave functions derived from high-precision phenomenological
 nuclear potentials, one-nucleon, and two-nucleon electroweak
 currents. 
 
 \subsection{Gamow-Teller operator}
 In contrast to the Fermi transition, the Gamow-Teller transition also involves two-body operators at NLO. The diagrams that contain a
 one-body weak interaction are coupled to $g_A$ and contain one ERE
 insertion up to NLO. The two-body diagrams are coupled to the two-body operators with prefactor $L_{1, A}$. By summing over all
 diagrams and comparing the resulting sum to the triton half-life,
 \cite{Chou}, $l_{1, A}$ can be extracted, as will be discussed later
 in Section~\ref{extraction}. We used the experimental input parameters shown in Tab. \ref{table_exp_data} for all numerical
 calculations.
 
 We emphasize furthermore that in the absence of Coulomb interaction,
 the Fermi transition matrix element at LO is 1 ({\it i.e.,}
 $\alpha=0$). Similarly, the LO matrix element of the Gamow-Teller
 transition with $\alpha=0$ was easily found to be:
 \begin{multline}
 \langle ||\text{GT}||\rangle^{\text{LO}}_{\alpha=0}=
 \dfrac{\langle\frac{1}{2}\|\boldsymbol{\sigma}\|\frac{1}{2}\rangle}{\sqrt{2}}{\Bigl\langle\frac{1}{2}\Bigl\|\tau^0\Bigr\|\frac{1}{2}\Bigr\rangle}\\\ \times
 \sum\limits_{\mu,\nu}y_\mu y_\nu
 {\psi^{^3\text{H}}_\mu(p')\otimes}\left[\delta_{\mu,\nu} \hat{\mathcal{I}}(0,0)
 +a_{\mu\nu} \hat{\mathcal{K}}(0,0)\right] \otimes{\psi^{^3\text{H}}_\nu(p)}\\
 =\dfrac{\sqrt{6}}{\sqrt{2}}=\sqrt{3}~,
 \end{multline}
 where $a_{\mu\nu}$ is given in \cref{eq_a_mu_nu_3H}. We performed this
 calculation in two ways: one with $\alpha=0$ for both the scattering
 amplitude and the matrix element, and the other with $\alpha= 0$ for
 the matrix element, but for different scattering lengths, similarly to
 the Fermi case. From Tab.~\ref{table_GT}, 
 it is clear that the bulk of the Coulomb effect originates from the
 strong isospin breaking, {\it i.e.,} different scattering lengths, and
 not from the explicit one-photon exchange diagrams. These results
 imply that for both the Fermi and Gamow-Teller transitions, the
 explicit Coulomb interaction, {\it i.e.,} one-photon exchange
 diagrams, can be calculated perturbatively since their contribution to
 the matrix element is very small compared to the isospin breaking
 effect. 
 
\begin{figure}[h!]
 \centerline{
 \includegraphics[width=1\linewidth]{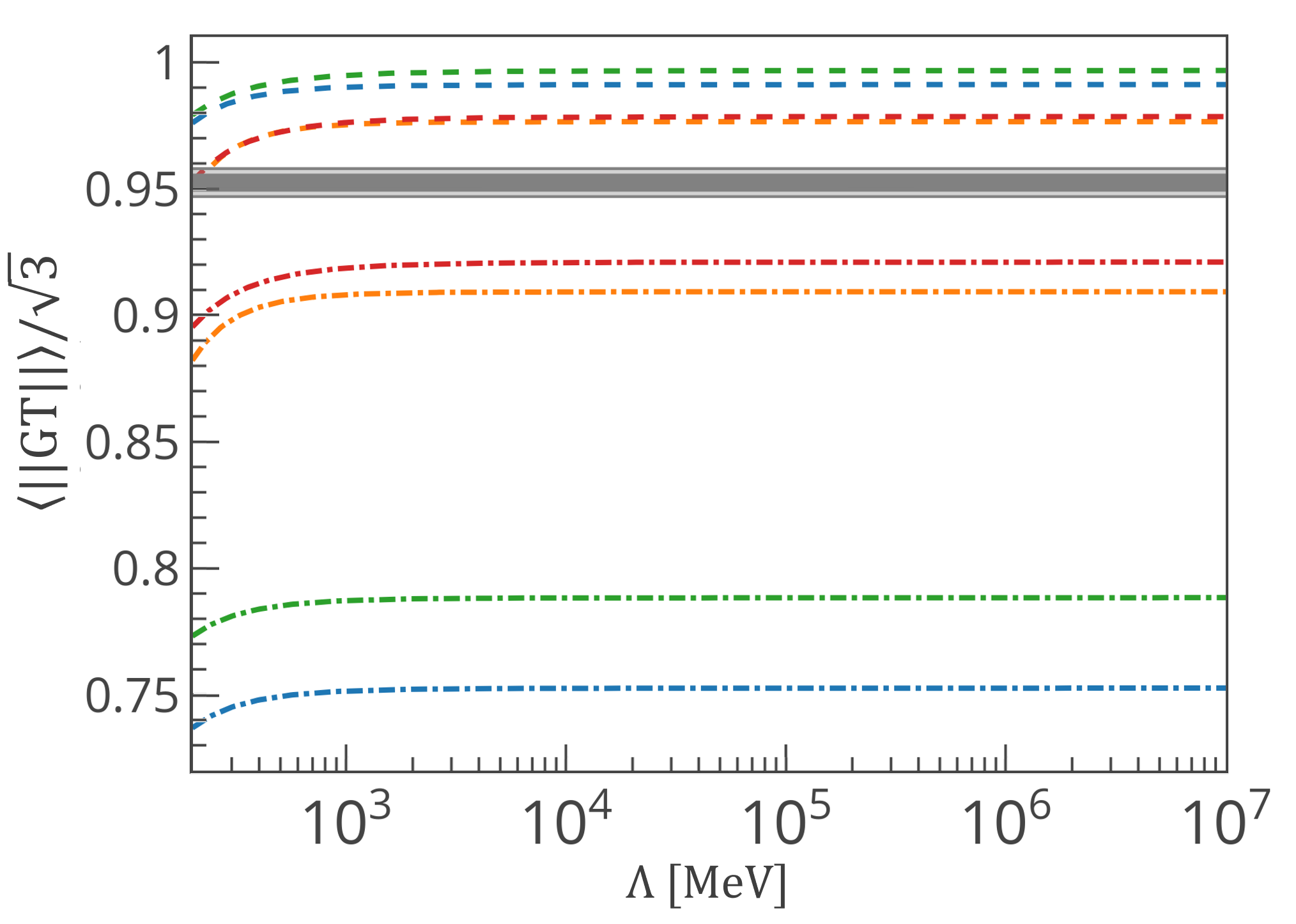}}
 \caption{\label{fig_full_GT}\footnotesize{ Numerical results of the Gamow-Teller
 transition. The gray area is the full $\langle ||\text{GT}|| \rangle$ matrix
 element with $g_A=1.273\pm 0.003\pm0.005$ \cite{gA_1.2701,
 doi:10.1063/1.4983578}. The blue (orange) short dashed line is the numerical
 result of $\langle ||\text{GT}|| \rangle^{\text{LO}}$ for the ERE- (Z-) parameterization, where $\alpha=0$ but $a_{np}\neq a_{pp}$.
 The green (red) short dashed line is the numerical
 result of $\langle ||\text{GT}|| \rangle^{\text{LO}}$ for the ERE- (Z-) parameterization, with $\alpha\neq0$.
 The blue (orange) dashed-dotted
 line is $\langle ||\text{GT}|| \rangle^{\text{NLO}}$ result with $l_{1, A}=0$ for the ERE- (Z-) parameterization, where $\alpha=0$ but $a_{np}\neq a_{pp}$.
 The green (red) dashed-dotted
 line is $\langle ||\text{GT}|| \rangle^{\text{NLO}}$ result with $l_{1, A}=0$ and $\alpha\neq0$, for the ERE- (Z-) parameterization.}}
 \end{figure}

 Our $\langle ||\text{GT}|| \rangle$ numerical results for both NLO arrangements
 are shown in Tab.~\ref{table_GT} and in Fig.~\ref{fig_full_GT}. The
 full NLO result with $l_{1, A}=0$ includes both one-body and two-body
 terms that contribute to $\langle ||\text{GT}|| \rangle$, without the diagrams
 that are coupled to $l_{1, A}$.
 
 \begin{table}[H]
 \begin{center}
 \begin{tabular}{l| c|c}
 \centering
 &$\langle ||\text{GT}|| \rangle$, ERE&$\langle ||\text{GT}|| \rangle$, $Z$\\
 \hline
 One-body, LO $\alpha=0$& $\sqrt{3}$&$\sqrt{3}$\\
 One-body, LO $\alpha=0$, $a_{np}\neq a_{pp}$&1.716&1.692 \\
 One-body, LO& 1.727& 1.695
 \\
 Full NLO, $l_{1, A}=0$, $\alpha=0$, $a_{np}\neq a_{pp}$&1.301&1.575\\
 Full NLO, $l_{1, A}=0$&1.383&1.596\\
 \end{tabular}
 \caption{\footnotesize{Numerical results for $\langle ||\text{GT}|| \rangle$. Note that the rows with the comment
 $``\alpha=0$, $a_{np}\neq a_{pp}$'' are without an explicit Coulomb force
 ($\alpha=0$) but with isospin breaking in the scattering
 lengths, {\it i.e.,} with physical values for the scattering
 lengths $a_{np} \neq a_{pp}$.}}
 \label{table_GT}
 \end{center}
 \end{table}
 
 \section{Extraction of the Gamow-Teller strength and fixing $l_{1, A}$}
 \label{extraction}
 The GT matrix element can be extracted from the triton half-life
 calculation using Eq.~\eqref{eq_GT}. The axial coupling constant,
 $g_A$, has been remeasured recently, leading to results whose range
 is much bigger than the current recommendation. To be on the safe
 side, we take $g_A=1.273\pm 0.003\pm0.005$
 \cite{doi:10.1063/1.4983578, gA_1.2701}. The first uncertainty in
 $g_A$ arises from the difference between the measurements of
 Refs.~\cite{gA_1.2701, doi:10.1063/1.4983578}, and the second
 uncertainty is the statistical experimental uncertainty. To extract
 the Gamow-Teller strength, we use our prediction for the Fermi
 transition: $\langle \text{F} \rangle=0.9993$
 \cite{fermi_reference_1998}. At large cutoff values, we find the
 empirical GT strength to be
 $\langle ||\text{GT}|| \rangle_\text{emp}=
 \sqrt{3}\frac{1.213\pm0.002}{g_A}$ \cite{fermi_reference_1998}. The
 uncertainty here originates mainly from the uncertainty in the triton
 half-life.
 
\begin{figure}[h!]
 \centerline{
 \includegraphics[width=1\linewidth]{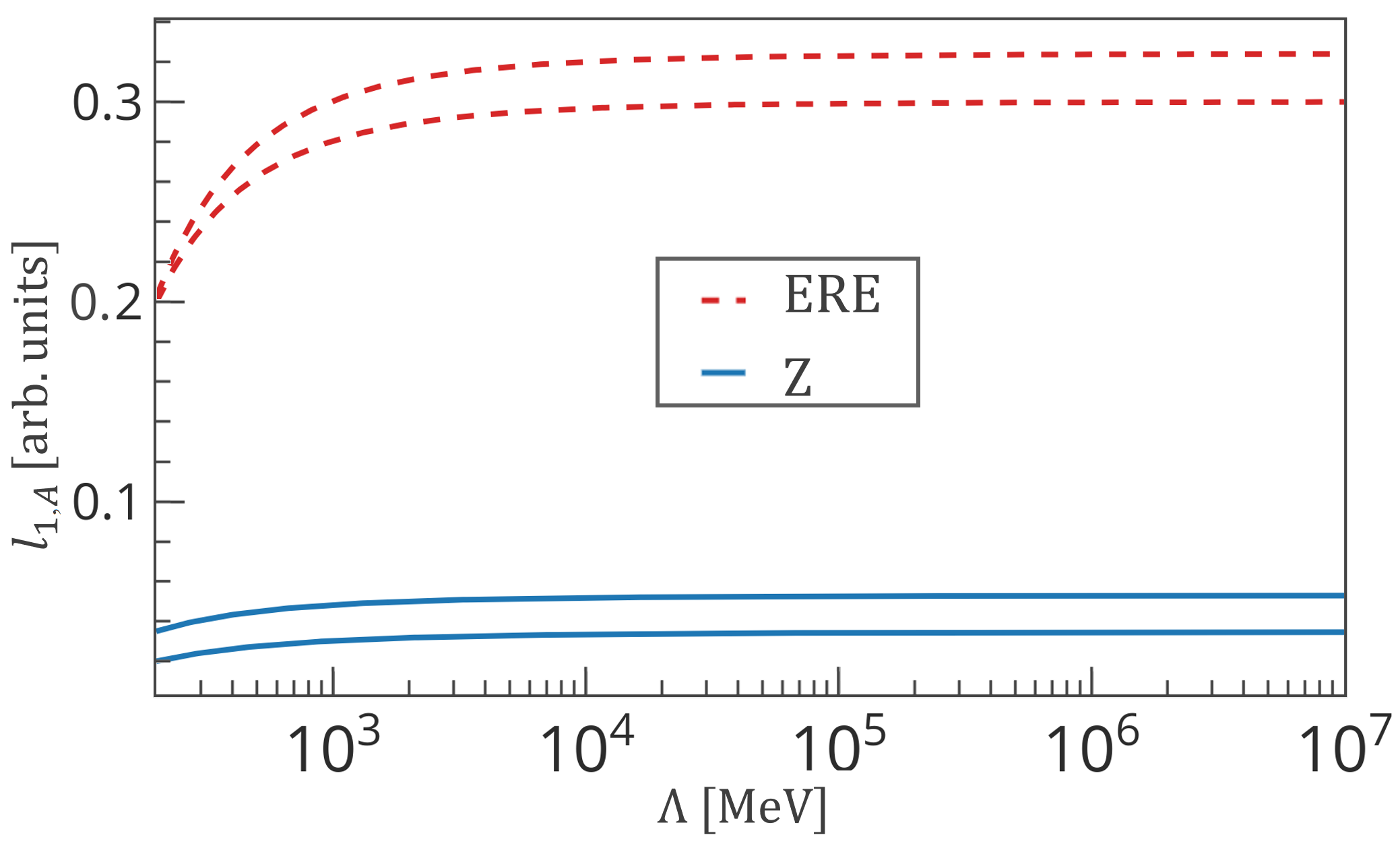}}
\caption{\label{fig_L1A} \footnotesize{Numerical results of
 $l_{1, A}$, with $g_A=1.273\pm 0.003\pm0.005$ \cite{gA_1.2701,
 doi:10.1063/1.4983578}. The dashed (solid) lines are the
upper and lower limits of the calculations in
 ERE- (Z-) parameterization.}}
 \end{figure}

 The difference between the empirical GT strength and the numerical
 result for the GT-transition at NLO is used to fix $l_{1, A}$ such that: 
 \begin{equation}\label{eq_L1A_emp}
 \l_{1,A}(\Lambda)= \frac{\langle||\text{GT}|| \rangle_{\text{emp}}-\langle||\text{GT}||\rangle_{l_{1, A}=0}^{\text{NLO}}}{\langle||\text{GT}||\rangle_{l_{1, A}}^{\text{NLO}} }~, 
 \end{equation}
 where $\langle ||\text{GT}|| \rangle_{l_{1, A}}^{\text{NLO}}$ are the two-body
 diagrams that contribute to the triton $\beta$-decay and are coupled
 to $l_{1, A}$, while $\langle ||\text{GT}|| \rangle_{l_{1, A}=0}^{\text{NLO}}$ is
 the sum over all the diagrams that contribute to the triton
 $\beta$-decay \textbf{without} the diagrams coupled to $l_{1, A}$.
 The numerical results for $l_{1,A}$ for both ERE- and
 $Z$-parameterizations are shown in Fig.~\ref{fig_L1A}.
 
 Importantly, we find numerically, that for both parameterizations,
 $l_{1,A}$ converges with increasing UV cutoff, a fact that has been
 already predicted by theory \cite{Butler:1999sv}, where:
 \begin{subequations}
 \begin{align}
 l_{1, A}^{\text{ERE}}
 &=0.312\pm0.004 \pm0.004\pm0.001\label{eq_L1A_ERE_1}\\ 
 l_{1, A}^{\text{Z}}
 &=0.051\pm0.004\pm0.004\pm0.001\label{eq_L1A_Zd_1}~,
 \end{align}
 \end{subequations}
 The first and second uncertainties come from the aforementioned
 difference between recent experimental determinations of $g_A$ and
 $g_A$ statistical uncertainties \cite{gA_1.2701,
 doi:10.1063/1.4983578} while the third uncertainty comes from the
 rest of the experimental uncertainties, such as the statistical
 uncertainties in the measured triton half-life.
 
\section{Previous extractions of $l_{1,A}$}
\label{previous_L1A}
Due to the importance of $l_{1, A}$, as the first two-body LEC
that appears in the pionless description of $pp$-fusion, its
determination has attracted much attention in the literature. In this
subsection, we review previous extractions of $l_{1, A}$ in the
\pilesseft and the latest predictions of the $pp$-fusion rate. 
 
Two main approaches were taken in previous studies to determine
$l_{1, A}$. In the first, an experimental value of a two-body weak
interaction process, usually at the cutoff $\mu=m_\pi$, was used for
matching. Among these reactions are the deuteron dissociation by
anti-neutrinos from reactors \cite{Butler:2002cw} and neutrino
reactions with the deuteron, as measured in SNO
\cite{PhysRevC.67.025801}.
Both references extract $L'_{1,A}$ as defined in
Eq.~\eqref{eq:L1Asingle-nucleon} as $4.0\pm 6.3$
(Ref.~\cite{Butler:2002cw}) and $3.6 \pm 4.6$
(Ref.~\cite{PhysRevC.67.025801}). Using Eq.~\eqref{eq:l1a-relation1}
and $\mu = m_\pi$ we can extract $l_{1, A} = 0.09 \pm 0.14$
and $l_{1, A}= 0.08 \pm 0.11$, respectively.

In both cases, the large uncertainties originate from statistical
errors in the experiments, due to the small cross-section for
neutrino-deuteron reactions. The authors of Ref.~\cite{L1A} proposed,
therefore, a precision measurement of muon capture on the deuteron,
with the aim of reducing the uncertainties by a factor of 3,
reflecting an estimated 2-3\% experimental uncertainty in the (then
proposed) ongoing MuSun experiment \cite{Andreev:2010wd}. It is
important to note that the $\mu^-d$ capture has a large energy
transfer, possibly too large for an application of \pilesseft. In all
these studies, the uncertainties are mainly experimental, due to the
uncertainty in the observable, {\it, i.e.,} neglecting the truncation
error.
 
A different approach was taken by Ando and collaborators in
Ref.~\cite{Ando_proton}. They used the hybrid calculation of the
$pp$-fusion rate from Park {\it et al.}~\cite{PhysRevC.67.055206}. The
authors took the ratio of the two-body strength over the one-body
strength was taken from this calculation and fixed $L_{1, A}$ to
reproduce this ratio in the \pilesseft regime. Ando {\it et al.}
defined the coefficient of the two-body counterterm as
$L_{1,A} = -\left( g_A \frac{\rho_s+\rho_t}{2 \sqrt{\rho_s
 \rho_t}}+\frac{l_{1,A}^{\rm Ando}}{M\sqrt{\rho_s \rho_t}}\right)$. Using
the value $l^{\rm Ando}_{1,A}=-0.5\pm 0.03$, we find
$l_{1,A}= 0.038\pm0.002$. The small uncertainty in this result is due
to the accurate triton half-life measurement that is used to fix the
undetermined counterterms in Ref.~\cite{PhysRevC.67.055206}. However,
this work has been criticized since it is not guaranteed that their
approach is consistent. It employs two very different models and
non-observable quantities to perform the matching.
 
In 2017, The Nuclear Physics with Lattice Quantum Chromo Dynamics
(NPLQCD) collaboration calibrated $L_{1, A}$ using the triton
$\beta$-decay \cite{PhysRevLett.119.062002} for the non-physical pion
mass $m_\pi= 805\mev$ and then extrapolated it to the physical pion
mass.

\section{Summary and outlook}
\label{summary}
In this paper, we have used the approach introduced in
Ref.~\cite{Big_paper} to study tritium $\beta$-decay in the framework
of pionless effective field theory at next-to-leading order (NLO).
The EFT approach used here is useful for robust and reliable
theoretical uncertainty estimates, based on neglected orders in the
EFT expansion. The results presented in this paper show that tritium
$\beta$-decay can be calculated reliably, that the corresponding
matrix elements are properly renormalized and that the
half-life displays, therefore, the necessary RG invariance at NLO.

We found furthermore that up to NLO, the Coulomb interaction {\it
 i.e.,} the one-photon exchange interaction, can be included
perturbatively in the calculation of this observable, which is
consistent with its effect on the $^3$He binding energy as already
discussed in Refs.~\cite{konig1,konig3,konig5, Big_paper}.
 
Tritium $\beta$-decay depends on two matrix elements, the Fermi
transition, which includes a one-body polar-vector part, and the
Gamow-Teller transition, which includes one- and two-body axial-vector
parts.
 
We have tested the correct renormalization of our perturbative
calculation of the Fermi and Gamow-Teller matrix elements with an
analysis of the residual cutoff dependence at cutoffs that are
significantly larger than the breakdown scale of the EFT.
 
We have used the NLO calculation to fix the NLO LEC, $l_{1, A}$ that
is needed for a high-accuracy prediction of the solar proton-proton
fusion rate \cite{Kong1, Kong2, Ando_proton}. The NLO correction that
originates from the $l_{1, A}$ counterterm (short-range corrections)
is about 3\% (15\%) of the tritium decay Gamow-Teller strength in the
Z-(ERE-) parameterization. The short-range corrections associated with
the ERE-parameterization are significantly larger than those
associated with the Z-parameterization, which implies that the
ERE-parameterization internal error is larger than that of the
Z-parameterization. Also, the fact that our calculation is carried
out within a consistent perturbative approach allows reliable
uncertainty estimates originating from the experimental uncertainties
in $g_A$ and the half-life measurement.
 
Knowledge of $l_{1, A}$ is required for higher-order calculations of a
number of weak processes, such as $pp$-fusion
\cite{solar2,Kong1,Kong2,Ando_proton} and muon capture
\cite{L1A,muon2,muon4,gazit09,Gazit:2008vm}. In the near future, we
intend to examine our result for $l_{1, A}$, and its uncertainty by
addressing these low-energy weak processes
\cite{PhysRevLett.110.192503, Acharya:2016kfl} when theoretical and
empirical uncertainties estimations must accompany this
prediction. Besides, the calibration of LEC from a $^3$H $\beta$-decay
for the prediction of a two-nucleon process (such as $pp$-fusion) is
based on the assumption that \pilesseft is the appropriate framework
for calculating observables in the $A=2$ and $A=3$ systems. However,
this consistency cannot be examined using the weak observables only
due to the small number of appropriate reactions. Hence, another set
of well-measured low-energy $A<4$ interactions with similar
characteristics to those of the weak reactions is needed for
validation and verification of \pilesseft. The strong analogy between
the {electromagnetic} to weak observables indicates that well-measured
electromagnetic observables can serve as the required candidates. We
intend to use our perturbative framework for calculating general
matrix element that can predict the low-energy
\textbf{electromagnetic} $A<4$ observables. These observables can
serve as a case study for estimating the theoretical uncertainty of
\pilesseft and LEC extractions from the $A<4$ observables predictions
\cite{magnetic_moments}.
 
\section*{Acknowledgment}
We thank Sebastian K$\ddot{\text{o}}$nig for the detailed comparison
of the $^3$He wave-functions. We also thank Jared Vanasse and Johannes
Kirscher, as well as the rest of the participants of the GSI-funded
EMMI RRTF workshop ER15-02: Systematic Treatment of the Coulomb
Interaction in Few-Body Systems, for valuable discussions, which have
contributed significantly to this work. The research of D.G and
H.D was supported by ARCHES and by the ISRAEL SCIENCE FOUNDATION
(grant No. 1446/16). The research of L.P was supported by the
National Science Foundation under Grant Nos. PHY-1516077 and
PHY-1555030, and by the Office of Nuclear Physics, U.S.~Department of
Energy under Contract No.\ DE-AC05-00OR2272.
 
\bibliography{references2}
\end{document}